\documentclass[preprint,showpacs,amsmath,amssymb,superscriptaddress,pre]{revtex4-1}
\usepackage{amssymb}
\usepackage{graphicx}
\usepackage[bf,SL,BF]{subfigure}
\usepackage{CJK}

\begin{document}
\title{Thermal rectification and negative differential thermal conductance in harmonic chains with nonlinear system-bath coupling}

\author{Yi Ming}
\email{meanyee@mail.ustc.edu.cn}
\affiliation{School of Physics and Material Science, Anhui University, Hefei, Anhui 230601, People's Republic of China}

\author{Hui-Min Li}
\affiliation{Supercomputing Center, University of Science and Technology of China, Hefei, Anhui 230026, People's Republic of China}

\author{Ze-Jun Ding}
\email{zjding@ustc.edu.cn}
\affiliation{Hefei National Laboratory for Physical Sciences at Microscale and Department of Physics, University of Science and Technology of China, Hefei, Anhui 230026, People's Republic of China}%

\date{\today}

\begin{abstract}
Thermal rectification and negative differential thermal conductance were realized in harmonic chains in this work. We used the generalized Caldeira-Leggett model to study the heat flow. In contrast to the most previous studies considering only the linear system-bath coupling, we considered the nonlinear system-bath coupling based on recent experiment [A. Eichler \emph{et al.}, Nat. Nanotech. \textbf{6}, 339 (2011)]. When the linear coupling constant is weak, the multiphonon processes induced by the nonlinear coupling allow more phonons transport across the system-bath interface and hence the heat current is enhanced. Consequently, thermal rectification and negative differential thermal conductance are achieved when the nonlinear couplings are asymmetric. However, when the linear coupling constant is strong, the umklapp processes dominate the multiphonon processes. Nonlinear coupling suppresses the heat current. Thermal rectification is also achieved. But the direction of rectification is reversed comparing to the results of weak linear coupling constant.
\end{abstract}
\maketitle

\section{Introduction}
In the past decade, phononics -- a science and engineering of manipulating heat -- has attracted intense interest from fundamental research as well as applied research\cite{Li2012, Pop2010, Marconnet2013, Balandin2012266, Maldovan2013, Roberts2011648}. There are two essential effects in phononics, thermal rectification and negative differential thermal conductance (sometimes referred to as negative differential thermal resistance). Thermal rectification allows heat current to flow preferably in one direction. The first nanoscale thermal rectifier was proposed theoretically in 2002 based on an one-dimensional (1D) nonlinear chain\cite{Terraneo02}. Since then, a variety of theoretical thermal rectifiers were proposed based on diverse nonlinear systems\cite{Li04, Li05, Hu06, Segal05, Segal05b, Segal06, Segal08, Wu09, Eckmann06, Casati07}. Inspired by the seminal experimental work demonstrating that heat current flow preferentially along the direction of decreasing mass density in asymmetrically mass-loaded nanotubes\cite{Chang06}, a lot of studies on thermal rectification were performed in nonlinear mass graded systems\cite{Yang07, Zeng08, Pereira2010, Pereira2011, Wang2012, Romero2014, Liu14} and asymmetric carbon based nanostructures\cite{Wu07, Yang08, Lee12, Bui2012} including asymmetric graphene nanoribbons\cite{Hu09, Wang12b, Wang14, Yang09, Zhang11, Zhong11, Zhong12}. All studies attribute thermal rectification to the intrinsic nonlinearity (anharmonicity) of the studied systems.

Negative differential thermal conductance refers to the effect that the heat current decreases as the applied temperature difference increases. This effect is the critical element to realize thermal transistors\cite{Li06}, thermal logic gates\cite{Wang07} and thermal memory\cite{Wang08}. Negative differential thermal conductance can be obtained in many nonlinear lattices\cite{Lan06, Li06, Yang07, He09, Shao09, Shao11, Zhong09, Zhong11b, Hu13, Pereira2010, Ai11b, Ai11, He10, Chan14}. Graphene nanoribbons are also the suitable platforms for practically realizing the negative differential thermal conductance\cite{Hu11, Ai12, Zhong12}. The intrinsic nonlinearities of the systems are the necessary conditions to achieve negative differential thermal conductance, although the interface resistance between two-segment nonlinear systems\cite{Lan06, Li06, He09, Zhong09, Zhong11b, Shao09, Shao11, Hu13} or the boundary resistance between heat baths and nonlinear systems\cite{Ai11, He10} are important.

However, the phonon mean-free path in graphene ($\sim 775$ nm near room temperature\cite{Balandin2011}) is much longer than the sizes of graphene nanoribbons. Therefore the intrinsic nonlinearity is insignificant and thus graphene nanoribbons can be regarded as harmonic systems. In harmonic systems, although exceptions exist\cite{Segal05, Segal05b, Segal06}, negative differential thermal conductance cannot be achieved and thermal rectification can only be obtained in quantum regime by asymmetric coupling with an additional self-consistent heat bath\cite{Ming10, Zhang10, Pereira2010a, *Pereira2011b, *Pereira2011c, Bandyopadhyay11, Ouyang2010, Xie2012}. However, the nanoscale self-consistent heat bath is hard to realize\cite{Bergfield2013}. Therefore, harmonic systems did not receive much interest in researches on thermal rectification and negative differential thermal conductance.

In almost all theoretical models mentioned above, the system-bath couplings were supposed to be linear. This is because the energy dissipations (dampings) in previous studied systems were supposed to be linear in general. However, recent researches have revealed that the nanostructures with high aspect ratio such as nanotubes and graphene nanoribbons can easily be driven into nonlinear dissipation regime\cite{Eichler2011}. As shown in Ref.~[\onlinecite{Eichler2011}], nonlinear dissipation can be treated as a generalized Caldeira-Leggett model with nonlinear system-bath coupling\cite{Barik05, Zaitsev12}. Nonlinear system-bath coupling corresponds to the inelastic boundary phonon scattering. In low-dimensional systems, thermal boundary conductance (also referred to as interfacial thermal conductance) becomes increasingly important\cite{Marconnet2013, Pop2010, Cahill03, *Cahill14}. At high temperature, many experimental\cite{Stoner92, *Stoner93, Stevens05, Lyeo06, Hopkins2007, Hopkins2008, Chen2009, Panzer10, Oh2011, Zhang2012, Norris2012, Duda2013, Dechaumphai14, Hohensee2015, Wang2015}, computational\cite{Twu2003, Chen2004, Stevens2004, Stevens2007, Hu2009, Landry2009, Ong2010, Duda2011, Chalopin2012, Pei2012, Khosravian2013, Saaskilahti2014, Liu2014, liu2014b, Zhang2015}, and empirical\cite{Duda2010, Duda2011b, Hopkins2009, Hopkins2009b, Hopkins2011} approaches have uncovered that the thermal boundary conductance at weakly bonded interface (or interface between highly dissimilar materials) exceeds the upper bound of elastic thermal conductance and nearly increases linearly with temperature. These results reveal that inelastic phonon scattering at interface contributes significantly to thermal boundary conductance. Therefore, nonlinear system-bath coupling is non-trivial for studying heat transport in low-dimensional systems.

In contrast to linear system-bath coupling, in thermal transport community, nonlinear system-bath coupling has received a little consideration in the previous works. In Refs.~[\onlinecite{Segal05, Segal05b, Segal06}], using small polaron transformation and based on master equation analysis, nonlinear system-bath coupling was considered and consequently the thermal rectification and negative differential thermal conductance were achieved in nonlinear two-level system and even in a harmonic molecular junction (a single harmonic oscillator). We noted that the nonlinear coupling is so strong that the linear coupling is omitted as shown in the Appendix of Ref.~[\onlinecite{Segal06}]. Therefore, the relative contributions of nonlinear coupling and linear coupling to thermal rectification and negative differential thermal conductance were not addressed.

In this work, we modeled the high-aspect-ratio nanostructure as an 1D harmonic chain. Two heat baths are coupled to it at the ends. The couplings are allowed to be nonlinear in addition to linear. Heat transport in the chain is studied at high temperature limit. When the linear system-bath couplings are weak, the numerical results reveal four effects of nonlinear system-bath coupling on heat current. Firstly, heat current is enhanced when nonlinear couplings are taken into account.  When the nonlinear coupling constant is weak, heat current is proportional to the square of the nonlinear coupling constant. Secondly, heat current increases linearly with the average temperature of the two baths when the temperature difference is fixed. When the nonlinear coupling constant is weak, the slope of increasing is proportional to the square of the nonlinear coupling constant. Thirdly, negative differential thermal conductance can be obtained in any temperature region by properly choosing the coupling constants. Lastly, thermal rectification is also obtained when the chain asymmetrically couples to the two baths. When both linear couplings are weak, the higher heat current is obtained when the hot bath couples to the chain with the stronger nonlinear coupling. All numerical results are consistent with our analytical results by approximately solving the generalized Langevin equation. When the linear system-bath couplings are strong, there is no available analytical result. Heat current is calculated numerically. Comparing with the results of weak coupling, heat current is suppressed by the nonlinear couplings. And heat current decreases with the average temperature of the two baths when the temperature difference is fixed. The slope of decreasing is also dependent on the nonlinear coupling constant. Moreover, the direction of thermal rectification is reversed. The higher heat current is obtained when the hot bath couples to the chain with the weaker nonlinear coupling. However, negative differential thermal conductance is not achieved in strong linear coupling regime.

The rest of the paper is organized as follows. In Sec. \ref{sec2}, the model and the analytical formulas and results are presented. The numerical results are presented in Sec. \ref{sec3}. Finally, we draw the conclusions and discuss the potential experimental realization of the nonlinear system-bath coupling as well as the range of validity of our approximate analytical results in Sec. \ref{sec4}.

\section{Model and methods}\label{sec2}
\subsection{Model}\label{sec2-1}
Harmonic chain contains $N$ particles is considered. These particles are connected by harmonic springs with equal spring constants which are chosen as equal to $1$. Then Hamiltonian of the chain is
\begin{equation}\label{chain}
H_S=\sum_{l=1}^N \frac{p_l^2}{2m_l}+\sum_{l=0}^{N+1} \frac{(x_l-x_{l+1})^2}{2},
\end{equation}
where $x_l$, $p_l$ and $m_l$ denote respectively the displacement of the $l$th particle from its equilibrium position, the momentum and the mass of the $l$th particle. The fixed boundary conditions are chosen as $x_0=x_{N+1}=0$. Two uncorrelated heat baths ($L$ and $R$) which are initially in thermal equilibrium at temperatures $T_L$ and $T_R$ are connected to the $1$st particle and the $N$th particle. Each bath is modeled by a collection of $M$ oscillators with harmonic interactions. The
Hamiltonian of each bath is
\begin{equation}\label{bath}
H_{B}=\sum_{\alpha=1}^M \frac{P_\alpha^2}{2}+\sum_{\alpha,\beta}\frac{1}{2}K_{\alpha\beta}Q_\alpha Q_\beta,
\end{equation}
where $Q_\alpha$ and $P_\alpha$ are the displacement and the momentum of the $\alpha$th unit-mass oscillator of the bath, $K_{\alpha\beta}$ is the spring constant between the $\alpha$th and the $\beta$th oscillator of the bath.

The $L$th ($R$th) oscillator of the left (right) bath is connected to the $1$st ($N$th) particle of the chain. The system-bath coupling Hamiltonian is
\begin{equation}\label{coup}
H_I=-g(x_1) Q_L-f(x_N) Q_R.
\end{equation}
Where $g(x_1)$ and $f(x_N)$ are functions of $x_1$ and $x_N$ for describing the coupling strength. If $g(x_1)$ (or $f(x_N)$) is proportional to the higher exponent of $x_1$ (or $x_N$) than $1$, the coupling is nonlinear in the coordinate of chain but linear in the heat-bath coordinates. Then the bath coordinates can be integrated out and we can obtain the generalized Langevin equations with multiplicative noises for the coordinates of chain. It should be mentioned that the coupling Hamiltonian in here is equal to those in Refs. [\onlinecite{Barik05}] and [\onlinecite{Zaitsev12}] by just transforming the heat-bath coordinate into its normal-mode coordinate space (as show in Appendix \ref{app0}). However, Hamiltonian (\ref{coup}) exhibits the direct meaning of coupling between two particles.

In the Markovian limit, the following generalized Langevin equations of the chain can be obtained according to the standard procedure\cite{Barik05, Zaitsev12, Dhar2003, Dhar2006, Dhar2008} as (see Appendix \ref{app0})
\begin{equation}\label{motion}
m_l \ddot{x}_l=-(2x_l-x_{l-1}-x_{l+1})-\gamma_l(x_l)\dot{x}_l+\xi_l(x_l).
\end{equation}
Where $\gamma_l(x_l)=\gamma_L[g^\prime(x_1)]^2\delta_{l,1}+\gamma_R[f^\prime(x_N)]^2\delta_{l,N}$ denotes dissipation and $\xi_l(x_l)=g^\prime(x_1) \eta_L \delta_{l,1}+f^\prime(x_N) \eta_R \delta_{l,N}$ is the noise term. The prime ($\prime$) indicates derivative with respect to the corresponding argument. At high temperature (classical limit), the fluctuation-dissipation relations $\langle\eta_L(t_1)\eta_L(t_2)\rangle_\eta=2k_B T_L \gamma_L \delta(t_2-t_1)$ and $\langle\eta_R(t_1)\eta_R(t_2)\rangle_\eta=2k_B T_R \gamma_R \delta(t_2-t_1)$ are satisfied. Where $\langle \cdots \rangle_\eta$ denotes an average over the noise.

To analytically study the heat current flowing in the chain, the Fokker-Planck equation corresponding to Eq. (\ref{motion}) is expressed as\cite{Reichl98}
\begin{eqnarray}\label{FP}
  && \frac{\partial P}{\partial t} = -\sum_{l=1}^N \frac{\partial}{\partial x_l} (v_l P)-\sum_{l=1}^N \frac{\partial}{\partial v_l} \left[\frac{-(2x_l-x_{l-1}-x_{x+1})}{m_l} P\right] \nonumber\\
  && +\frac{\gamma_L}{m_1}[g^\prime (x_1)]^2 \frac{\partial}{\partial v_1} (v_1 P)+\frac{\gamma_L k_B T_L}{m_1^2} [g^\prime (x_1)]^2 \frac{\partial^2}{\partial v_1^2} P \nonumber\\
  && +\frac{\gamma_R}{m_N}[f^\prime (x_N)]^2 \frac{\partial}{\partial v_N} (v_N P)+\frac{\gamma_R k_B T_R}{m_N^2} [f^\prime (x_N)]^2 \frac{\partial^2}{\partial v_N^2} P.
\end{eqnarray}
Where $v_l=p_l/m_l=\dot{x}_l$ is the velocity of the $l$th particle of the chain. $P=P(\{x_l\},\{v_l\},t)$ is the phase-space probability density function. From the Fokker-Planck equation (\ref{FP}), the time derivative for the energy of the chain $\partial \langle H_S \rangle/\partial t$ can be calculated. Where $\langle \cdots \rangle$ implies an ensemble average over the whole phase space of the chain. Then the heat current in steady state with $\partial \langle H_S \rangle_{st}/\partial t =0$ can be defined via the continuity equation. For simplicity, in this work, we choose the coupling functions as polynomial in $x_1$ and $x_N$ only up to the quadratic terms as in Ref.~[\onlinecite{Zaitsev12}]: $g(x_1)=k_L x_1+\mu_L x_1^2/2$ and $f(x_N)=k_R x_N+ \mu_R x_N^2/2$. Where $k_L$ and $k_R$ are linear coupling constants. $\mu_L$ and $\mu_R$ are nonlinear coupling constants. As a consequence, the steady-state heat current is defined as
\begin{equation}\label{current}
J^{st}=\frac{1}{2}(J_L^{st}-J_R^{st})
\end{equation}
with
\begin{eqnarray}\label{heat}
&&J_{L(R)}^{st}=\frac{\gamma_{L(R)} k_B T_{L(R)}}{m_{1(N)}}\Big(k_{L(R)}^2+\mu_{L(R)}^2 \big\langle x_{1(N)}^2 \big\rangle\Big)\nonumber\\
&&-\gamma_{L(R)} \Big(k_{L(R)}^2 \big\langle v_{1(N)}^2 \big\rangle+\mu_{L(R)}^2 \big\langle x_{1(N)}^2 v_{1(N)}^2 \big\rangle \Big).
\end{eqnarray}
Where the subscripts without brackets corresponding to the heat current flowing into the chain from the left bath and the subscripts in brackets corresponding to the heat current flowing into the chain from the right bath.

\subsection{Perturbation approximation}\label{sec2-2}
For nonlinear system-bath coupling, there are no rigorous results about $J^{st}$ even for harmonic chain. However, when the couplings are linear, heat current flowing through harmonic chain can be obtained exactly as shown in Refs.~[\onlinecite{Dhar2001, Dhar2003, Dhar2006, Dhar2008}]. Therefore, approximate analytical results can be obtained by using a proper perturbation scheme when the nonlinear couplings are weak. Only considering the linear system-bath coupling by letting $\mu_L=\mu_R=0$ as the zeroth approximation, $x_{1(N)}$ and $v_{1(N)}$ can be calculated for harmonic chain via the equations of motion (\ref{motion})\cite{Dhar2001, Dhar2003, Dhar2006, Dhar2008}. Then the heat current can be obtained by inserting these zeroth approximations of $x_{1(N)}$ and $v_{1(N)}$ into Eqs. (\ref{current}) and (\ref{heat}).

In the linear coupling approximation with $\mu_L=\mu_R=0$, following Refs.~[\onlinecite{Dhar2001, Dhar2003, Dhar2006, Dhar2008}], the equations of motion (\ref{motion}) can be solved exactly by taking the Fourier transformation. The results can be expressed as
\begin{eqnarray}\label{app-eqn1}
x_l(t)&=&\frac{1}{2\pi}\int_{-\infty}^\infty d\omega \hat{Z}_{lm}(\omega) \hat{\xi}_m(\omega) e^{i\omega t},\quad \text{where}\nonumber\\
\hat{Z}&=&\hat{Y}^{-1} \qquad\qquad\qquad\qquad\qquad\qquad \text{with}\nonumber\\
\hat{Y}&=&\hat{\Phi}-\omega^2\hat{M}-\hat{\Gamma}(\omega),\nonumber\\
\hat{\Phi}_{lm}&=&-\delta_{l,m+1}+2\delta_{l,m}-\delta_{l,m-1},\nonumber\\
\hat{M}_{lm}&=&m_l\delta_{l,m},\nonumber\\
\hat{\Gamma}_{lm}&=&\delta_{l,m}[-i \gamma_L k_L^2 \omega \delta_{l,1}-i \gamma_R k_R^2 \omega \delta_{l,N}],\nonumber\\
\hat{\xi}_l&=&\eta_L(\omega) k_L \delta_{l,1}+\eta_R(\omega) k_R \delta_{l,N}.
\end{eqnarray}
Letters with hat symbol represent the matrices. The superscript $-1$ means the inversion of the corresponding matrix. The corresponding fluctuation-dissipation relations in Fourier space are $\langle \eta_{L(R)}(\omega_1) \eta_{L(R)}(\omega_2) \rangle_\eta=4\pi \gamma_{L(R)} k_B T_{L(R)} \delta(\omega_1+\omega_2)$.

Substituting Eq.~(\ref{app-eqn1}) into Eq.~(\ref{heat}), the results can be expressed as
\begin{widetext}
\begin{eqnarray}\label{app-eqn2}
  J_L^{st} &=& \frac{\gamma_L k_L^2 k_B T_L}{m_1}-\gamma_L k_L^2 \langle v_1^2 \rangle +\mu_L^2 \left( \frac{\gamma_L k_B T_L}{m_1} \langle x_1^2 \rangle -\gamma_L \langle x_1^2 v_1^2 \rangle\right) \nonumber\\
  &=& \frac{\gamma_L k_L^2 \gamma_R k_R^2 k_B}{\pi}(T_L-T_R) \int_{-\infty}^\infty d\omega \omega^2 \hat{Z}_{1N}(\omega) \hat{Z}_{1N}(-\omega) \nonumber\\
  &&+ \mu_L^2 \frac{2 \gamma_L}{\pi^2} \left[\gamma_L k_L^2 k_B T_L \int_{-\infty}^\infty d\omega \omega \hat{Z}_{11}(\omega) \hat{Z}_{11}(-\omega) + \gamma_R k_R^2 k_B T_R \int_{-\infty}^\infty d\omega \omega \hat{Z}_{1N}(\omega) \hat{Z}_{1N}(-\omega) \right]^2\nonumber\\
  && + \left[\frac{\gamma_L k_L^2 \gamma_R k_R^2 k_B}{\pi}(T_L-T_R) \int_{-\infty}^\infty d\omega \omega^2 \hat{Z}_{1N}(\omega) \hat{Z}_{1N}(-\omega)\right] \nonumber\\
  &&\times\frac{\mu_L^2}{\pi k_L^2} \left[\gamma_L k_L^2 k_B T_L \int_{-\infty}^\infty d\omega \hat{Z}_{11}(\omega) \hat{Z}_{11}(-\omega) + \gamma_R k_R^2 k_B T_R \int_{-\infty}^\infty d\omega \hat{Z}_{1N}(\omega) \hat{Z}_{1N}(-\omega) \right].
\end{eqnarray}
\end{widetext}
Where
\begin{equation}\label{app-eqn3}
\hat{M}^{-1}=\frac{1}{\pi}\int_{-\infty}^\infty d\omega \hat{Z}(\omega) \omega^2 \hat{\Gamma} \hat{Z}(-\omega)
\end{equation}
is used to obtain the second equality\cite{Casher71, Ishii1973}. We should mention that Eq.~(\ref{app-eqn3}) is satisfied only when $T_L=T_R$. Therefore, Eqs.~(\ref{app-eqn4}), (\ref{app-eqn5}) and (\ref{rect}) are approximate results for $T_L\approx T_R$. Furthermore, the second term of the last equality equals to zero because $\omega \hat{Z}_{11}(\omega) \hat{Z}_{11}(-\omega)$ and $\omega \hat{Z}_{1N}(\omega) \hat{Z}_{1N} (-\omega)$ are odd functions of $\omega$. Then Eq.~(\ref{app-eqn2}) can be simplified as
\begin{widetext}
\begin{eqnarray}\label{app-eqn4}
  &&J_L^{st} = \kappa_0 (T_L-T_R) \left[ 1+\frac{\mu_L^2}{\pi k_L^2} (\gamma_L k_L^2 k_B T_L I_{11} + \gamma_R k_R^2 k_B T_R I_{1N} ) \right].
\end{eqnarray}
Similarly, one can obtain
\begin{eqnarray}\label{app-eqn5}
  &&J_R^{st} = \kappa_0 (T_R-T_L) \left[ 1+\frac{\mu_R^2}{\pi k_R^2} (\gamma_R k_R^2 k_B T_R I_{NN} + \gamma_L k_L^2 k_B T_L I_{N1}) \right] .
\end{eqnarray}
Where $\kappa_0=\gamma_L k_L^2 \gamma_R k_R^2 k_B K_{1N}/\pi$, $K_{lm}=\int_{-\infty}^\infty d\omega \omega^2 \hat{Z}_{lm}(\omega) \hat{Z}_{lm}(-\omega)$ and $I_{lm}=\int_{-\infty}^\infty d\omega \hat{Z}_{lm}(\omega) \hat{Z}_{lm}(-\omega)$. $\kappa_0 (T_L-T_R)$ is just the heat current obtained in Refs.~[\onlinecite{Dhar2001, Dhar2003, Dhar2006, Dhar2008}] for linear system-bath coupling. Which is valid for high temperature difference. Therefore, in spite of the approximation (\ref{app-eqn3}) is used, our results are supposed to valid also for high temperature difference.
\end{widetext}

As shown in Ref.~[\onlinecite{Segal06}], the temperature difference $\Delta T$ can be imposed in two different ways. Firstly, we set
\begin{equation}\label{caseA}
(A)\quad T_L=T_0+\Delta T/2,\qquad T_R=T_0-\Delta T/2.
\end{equation}
Then the heat current (\ref{current}) can be expressed as
\begin{widetext}
\begin{eqnarray}\label{rect}
  J^{st} &=& \kappa_0\left\{1+\frac{1}{2\pi} \left[(\mu_L^2\gamma_L I_{11}+\mu_R^2\gamma_R I_{NN}) +(\frac{k_R^2}{k_L^2} \mu_L^2\gamma_R +\frac{k_L^2}{k_R^2}\mu_R^2\gamma_L) I_{1N} \right] k_B T_0 \right\}\Delta T \nonumber \\
  &&+\frac{1}{4\pi} \kappa_0 \left[(\mu_L^2\gamma_L I_{11}-\mu_R^2\gamma_R I_{NN})+(\frac{k_L^2}{k_R^2}\mu_R^2\gamma_L -\frac{k_R^2}{k_L^2}\mu_L^2\gamma_R) I_{1N}\right] k_B(\Delta T)^2.
\end{eqnarray}
\end{widetext}

When the system-bath couplings are linear with $\mu_L=\mu_R=0$, the heat current $J^{st}=\kappa_0 \Delta T$ depends linearly on the temperature difference $\Delta T$ but is independent on $T_0$ as shown in Eq.~(\ref{rect}). However, when the couplings are nonlinear (e.g., $\mu_L=\mu_R=\mu \neq 0$, $\gamma_L=\gamma_R=\gamma$ and $k_L=k_R=k$, then $J^{st}=\kappa_0 \Delta T +\kappa_0\mu^2\gamma(I_{11}+I_{1N})k_B T_0 \Delta T/\pi$), heat current is enhanced relative to $\kappa_0 \Delta T$ and it increases linearly with $T_0$ because $I_{lm}\ge 0$ (as shown in Appendix \ref{app2}, $\hat{Z}_{lm} (-\omega)=\hat{Z}^*_{lm} (\omega)$). These results are consistent with the results in Refs.~[\onlinecite{Stoner92, *Stoner93, Stevens05, Lyeo06, Hopkins2007, Hopkins2008, Chen2009, Panzer10, Oh2011, Zhang2012, Norris2012, Duda2013, Dechaumphai14, Hohensee2015, Wang2015, Twu2003, Chen2004, Stevens2004, Stevens2007, Hu2009, Landry2009, Ong2010, Duda2011, Chalopin2012, Pei2012, Khosravian2013, Saaskilahti2014, Liu2014, liu2014b, Zhang2015, Duda2010, Duda2011b, Hopkins2009, Hopkins2009b, Hopkins2011}].

As one can expect, thermal rectification is absent when the couplings are linear with $\mu_L=\mu_R=0$. But the situation becomes very different when the system-bath couplings are nonlinear. Thermal rectification is achieved when the second term of Eq.~(\ref{rect}) is nonzero. If the system-bath couplings are symmetric with $k_L= k_R$, $\gamma_L=\gamma_R$ and $\mu_L=\mu_R$, thermal rectification is absent. However, if the couplings are asymmetric, thermal rectification can be achieved. These approximate analytical results are verified by the following numerical results.

For the second way to impose the temperature difference, we set
\begin{equation}\label{caseB}
(B)\quad T_L=T_s,\qquad T_R=T_s-\Delta T.
\end{equation}
Where $\Delta T \le T_s$ to ensure $T_R\ge 0$. The heat current (\ref{current}) is thus expressed as
\begin{widetext}
\begin{eqnarray}\label{NDTC1}
  J^{st} &=& \kappa_0\left\{1+\frac{1}{2\pi} \left[(\mu_L^2\gamma_L I_{11}+\mu_R^2\gamma_R I_{NN}) +(\frac{k_R^2}{k_L^2} \mu_L^2\gamma_R +\frac{k_L^2}{k_R^2} \mu_R^2\gamma_L) I_{1N} \right] k_B T_s \right\}\Delta T \nonumber \\
  &&-\frac{1}{2\pi} \kappa_0 (\mu_R^2 I_{NN}+\frac{k_R^2}{k_L^2}\mu_L^2 I_{1N})\gamma_R k_B(\Delta T)^2.
\end{eqnarray}
\end{widetext}
 It indicates that, when $\Delta T>0$, heat current first increases with $\Delta T$, and then decreases after reaching a maximum. From Eq.~(\ref{NDTC1}), the temperature difference corresponding to the maximum heat current is
 \begin{equation}\label{NDTC2}
 (\Delta T)_m=\frac{T_s}{2}+\frac{2\pi +(\mu_L^2 I_{11}+\frac{k_L^2}{k_R^2}\mu_R^2 I_{1N})\gamma_L k_B T_s}{2(\mu_R^2 I_{NN}+\frac{k_R^2}{k_L^2}\mu_L^2 I_{1N})\gamma_R k_B}>\frac{T_s}{2}.
 \end{equation}
 When letting $k_R/k_L\gg 1$, by choosing the suitable $\mu_L$ and $\mu_R$, one can expect that $(\Delta T)_m<T_s$ from Eq. (\ref{NDTC2}) and thus $T_R>0$. Therefore, negative differential thermal conductance occurs with the onset temperature difference being $(\Delta T)_m$. One should note that $T_R$ has to less than $T_s/2$ to achieve the negative differential thermal conductance. However, there is no limitation on temperature region to realize negative differential thermal conductance. In contrast to the results of Ref.~[\onlinecite{Segal06}], if the system-bath couplings are symmetric, the temperature difference calculated from Eq.~(\ref{NDTC2}) is larger than $T_s$ and thus negative differential thermal conductance cannot be achieved.

\section{Numerical results}\label{sec3}

To obtain the heat current in harmonic chains, we use the implicit midpoint algorithm\cite{Burrage07} to integrate the equations of motion (\ref{motion}). (The results have been compared with those obtained by using Mannella's leapfrog algorithm\cite{Burrage07} and the velocity Verlet algorithm\cite{Allen87}. The differences are negligible.) Equilibration times ranged from $10^8$ - $10^9$ time steps of step size $0.05$ and steady-state averages were taken over another $10^8$ - $10^9$ time steps. (The results have been compared with those obtained by setting time step size as $0.005$. The differences are negligible.) The steady state is reached by checking whether the results of different equilibration times are equal and checking whether the local heat currents are constant along the chain. In all simulations, we study the equal-mass harmonic chains with $m_l=1$. Moreover, we set $k_B=1$ and $N=24$. Therefore, the cut-off frequency of the harmonic chain is $2$. Comparing with the cut-off frequency of the out-of plane acoustic (ZA) phonon polarization branches of graphene ($\sim 14$ THz\cite{Balandin2011}), the real temperature $T_{real}$ is related to the dimensionless temperature $T$ through the relation $T_{real}\approx 336\, T$ (K). In this work, $T$ is chosen in the range from $1$ to $2$, thus the corresponding $T_{real}$ is in the range from $336$ K to $672$ K.

The local heat current at site $l$ is defined as $J_l=\langle (\dot{x}_l+\dot{x}_{l-1})f_{l,l-1} \rangle/2$, where $f_{l,l-1}$ is the force exerted by the $(l-1)$th particle on the $l$th particle and $\langle \cdots\rangle$ denotes a steady-state average. At steady state, $J_l$ is independent on the site position $l$. In our simulations, the heat current flows from the left bath to the right bath is defined as $J_{+}=\sum_{l=2}^N J_l/(N-1)$. Reversing the temperature difference, the heat current flows in the reverse direction is denoted as $J_-$.

\subsection{Weak linear coupling constant}
\begin{center}
\begin{figure}[htbp]
\includegraphics[width=0.8\textwidth]{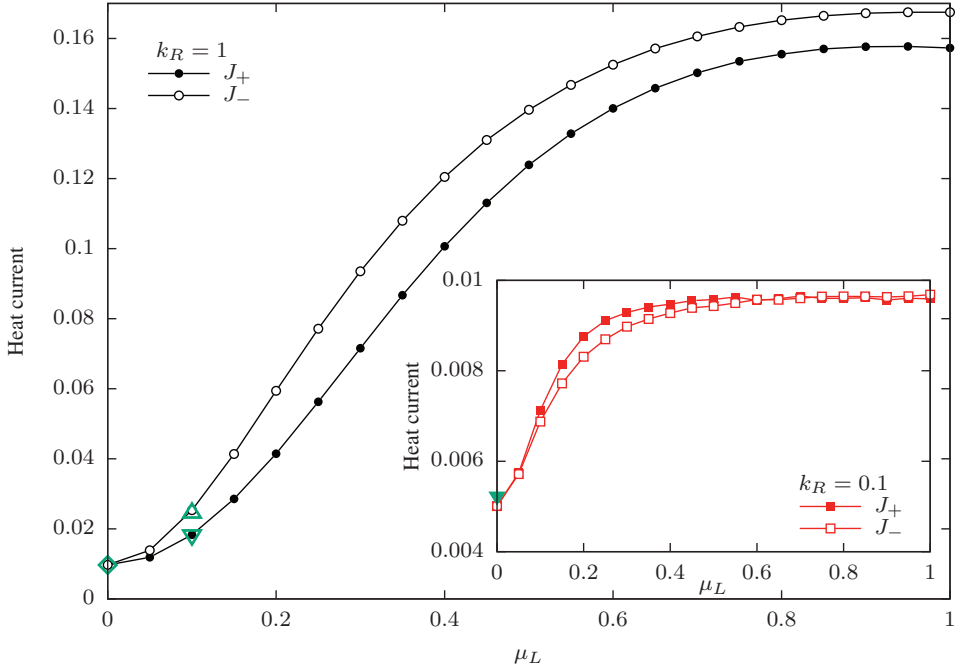}
\caption{\label{fig1} Heat currents $J_{+}$ and $J_{-}$ as functions of $\mu_L$. The temperature difference is imposed as model A (\ref{caseA}). The parameters are $T_0=1.5$, $\Delta T=1$ for $J_+$ and $\Delta T=-1$ for $J_-$, $\gamma_L=\gamma_R=1$, $\mu_R=0$, $k_L=0.1$ and $k_R=1$. The open diamond corresponds to $J_-$ for $\mu_L=1$ in the inset. The open up-triangle and the open down-triangle correspond to $J_+$ and $J_-$ for $\mu_L=1$ in Fig.~\ref{fig2}. In the inset, none parameter is changed but $k_R=0.1$. The solid down-triangle corresponds to the intercept $d_1$ in Fig.~\ref{fig3}.}
\end{figure}
\end{center}

We set $\mu_R=0$ in Fig.~\ref{fig1} and the inset. Only the left system-bath coupling is nonlinear. The linear coupling constants are set as $k_L=0.1$ and $k_R=1$ in Fig.~\ref{fig1}. As revealed in Ref.~[\onlinecite{Zhang2011}], without the nonlinear coupling, phonons in the whole frequency domain can transport across the right system-bath interface with the transmission equals to one. However, only the low-frequency phonons can transport across the left system-bath interface. If the left system-bath coupling is nonlinear, high-frequency phonons can transport across the left interface via the multiphonon process and thus the heat current is enhanced just as depicted in Fig.~\ref{fig1} and the inset. The enhancement is consistent with our analytical results in Sec.~\ref{sec2-2} and the results of weak coupling in Refs.~[\onlinecite{Landry2009, Chen2009, Hopkins2009b, Duda2013, Norris2012, Duda2010, Hohensee2015, Hopkins2011, Saaskilahti2014, Stoner92, *Stoner93, Stevens2004, Lyeo06, Stevens05}]. Moreover, $J_+$ and $J_-$ are proportional to the square of $\mu_L$ in Fig.~\ref{fig1} when $\mu_L$ is small as predicted in Eq.~(\ref{rect}).

Thermal rectification is apparent in Fig.~\ref{fig1}. Heat current flow preferably from the right bath to the left bath (i.e. $J_->J_+$) when $\mu_L\neq 0$. This is because the right system-bath interface is transparent to phonons. If the right bath is hotter, more phonons can be excited in the chain to participate in the multiphonon processes. Therefore, thermal rectification with $J_->J_+$ is obtained. With the increasing of nonlinear coupling constant $\mu_L$, the transmission of phonons across the left interface approach saturation. Consequently, the heat currents saturate as shown in Fig.~\ref{fig1}. However, when the right coupling is weak with $k_R=0.1$, thermal rectification is not evident. $J_+$ is only little higher than $J_-$ when $\mu_L<0.6$ as shown in the inset of Fig.~\ref{fig1}. This is because $k_R=0.1$ allows only low-frequency phonons transport across the right interface. Benefiting by the left nonlinear system-bath coupling, more low-frequency phonons can be excited in harmonic chain. Therefore, $J_+$ is little higher than $J_-$ when $\mu_L<0.6$. When $\mu_L>0.6$, the heat currents ($J_+$ and $J_-$) approach the saturated value. This saturated value at $\mu_L=1$ is marked as an open diamond point in Fig.~\ref{fig1} by corresponding the right interface in the inset to the left interface in Fig.~\ref{fig1}. The agreement reveals the fact that the transmissions of low-frequency phonons across the left interface approach one when $\mu_L>0.6$ in the inset.
\begin{center}
\begin{figure}[htbp]
\includegraphics[width=0.8\textwidth]{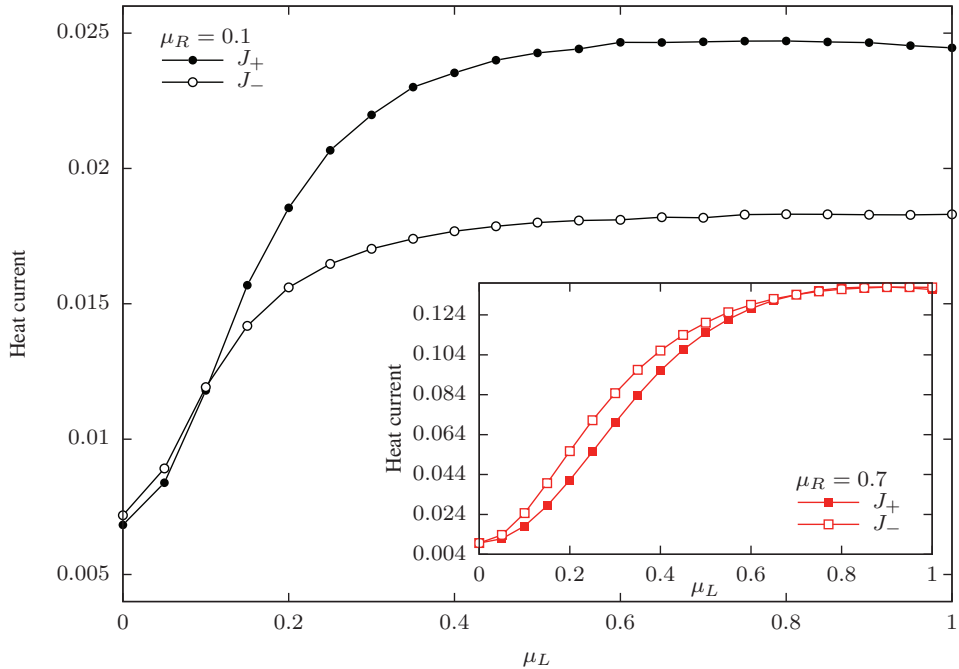}
\caption{\label{fig2} Heat currents $J_{+}$ and $J_{-}$ as functions of $\mu_L$. The temperature difference is imposed as model A (\ref{caseA}). The parameters are $T_0=1.5$, $\Delta T=1$ for $J_+$ and $\Delta T=-1$ for $J_-$, $\gamma_L=\gamma_R=1$, $\mu_R=0.1$ and $k_L=k_R=0.1$. In the inset, none parameter is changed but $\mu_R=0.7$.}
\end{figure}
\end{center}

Heat current is further enhanced when the right coupling is nonlinear. Comparing with the inset of Fig.~\ref{fig1}, we set $\mu_R=0.1$ in Fig.~\ref{fig2}. Thermal rectification is apparent with $J_+/J_-\approx 1.336$ at $\mu_L=1$. In addition, the direction of thermal rectification is reversed at $\mu_L=\mu_R=0.1$. This reversing of thermal rectification indicates that heat current is higher when the hot bath is coupled to the chain with stronger nonlinear coupling constant. It is consistent with the results of nonlinear coupling in Ref.~[\onlinecite{Segal06}]. This is based on the aforementioned mechanism that the stronger the nonlinear coupling is, the more the phonons can be excited in the chain to participate in the multiphonon process. The saturated values of $J_+$ and $J_-$ at $\mu_L=1$ are marked as the open up-triangle and the open down-triangle in Fig.~\ref{fig1} by corresponding the right interface in Fig.~\ref{fig2} to the left interface in Fig.~\ref{fig1}. We can find that the corresponding values are equal. This indicates that the left interface is transparent for all the phonons which can transport across the right interface when $\mu_L>0.6$. Based on this mechanism, thermal rectification will absent (i.e. $J_+=J_-$) when both $\mu_L$ and $\mu_R$ are higher than $0.6$. This is confirmed in the inset of Fig.~\ref{fig2} with $\mu_R=0.7$. Moreover, in the inset, $J_->J_+$ when $\mu_L<0.6$ because $\mu_R>\mu_L$.
\begin{center}
\begin{figure}[htbp]
\includegraphics[width=0.8\textwidth]{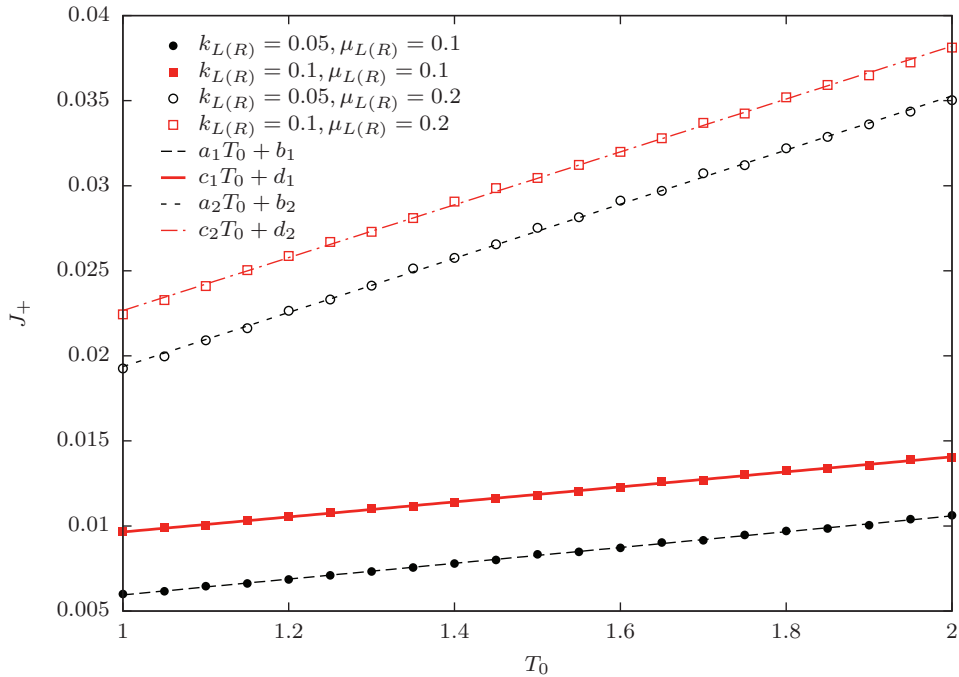}
\caption{\label{fig3} Heat currents $J_{+}$ as functions of the average temperature $T_0$. The temperature difference is imposed as model A (\ref{caseA}). The parameters are $\Delta T=1$ and $\gamma_L=\gamma_R=1$. The solid symbols and the open symbols correspond to $\mu_L=\mu_R=0.1$ and $\mu_L=\mu_R=0.2$. The square symbols and the circle symbols correspond to $k_L=k_R=0.1$ and $k_L=k_R=0.05$. The long-dash, solid, short-dash and dot-dash lines are linear fits of the data. The fitting parameters are $a_1=0.00464141$, $b_1=0.00130853$, $c_1=0.00441571$, $d_1=0.0052321$, $a_2=0.0159123$, $b_2=0.00345535$, $c_2=0.0155388$ and $d_2=0.00712554$.}
\end{figure}
\end{center}

In the high temperature limit, the phonon population in heat bath increases linearly with temperature. As a consequence, the heat conductivity increases with temperature when the system-bath coupling is nonlinear. The heat conductivity is defined as $\lim\limits_{\Delta T\rightarrow 0} J_{st}/\Delta T$. When the system-bath couplings are symmetric with $\mu_L=\mu_R$ and $k_L=k_R$, heat current $J_{st}$ is proportional to the temperature difference $\Delta T$. Therefore, fixing $\Delta T$, heat current $J_{st}$ increases linearly with the average temperature $T_0$, which is predicted by Eq.~(\ref{rect}). The numerical results in Fig~\ref{fig3} confirm this prediction. Moreover, $a_1\approx c_1$ and $a_2\approx c_2$ are obtained. Which indicates that the slopes of the linear fits are $\mu_{L(R)}$ dependent when $k_L=k_R$. The same $\mu_{L(R)}$ corresponds to the same slope. The higher the $\mu_{L(R)}$ is, the higher the slope is. This is consistent with Eq.~(\ref{rect}) and Ref.~[\onlinecite{Duda2011}]. As predicted in Eq.~(\ref{rect}), when the system-bath couplings are symmetric, the slope is proportional to $\mu_L^2$ and the intercept equals to $\kappa_0\Delta T$. The intercept $d_1$ is plotted in the inset of Fig.~\ref{fig1} as a solid down-triangle point. It approaches the predicted value $\kappa_0\Delta T$ for $k_L=k_R=0.1$. However, the intercept $d_2$ corresponding to $\mu_L=\mu_R=0.2$ is larger than $\kappa_0\Delta T$ for $k_L=k_R=0.1$. In addition, the ratios of the slopes are $a_2/a_1\approx 3.43$ and $c_2/c_1\approx 3.52$. They are smaller than the corresponding ratio of $\mu_L^2$, which is $(0.2/0.1)^2=4$. Besides the fitting errors and the numerical errors, we attribute these discrepancies to the fact that the approximation in Eq.~(\ref{rect}) is crude and it is only valid for small $\mu_{L(R)}$.
\begin{center}
\begin{figure}[htbp]
\includegraphics[width=0.8\textwidth]{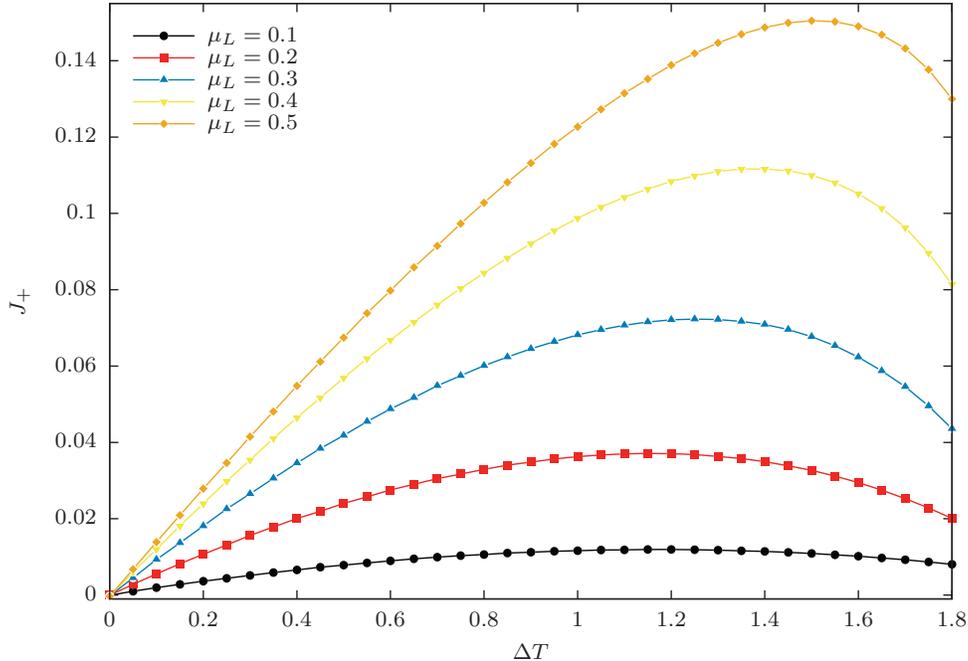}
\caption{\label{fig4} Heat current $J_{+}$ as function of $\Delta T$. The temperature difference is imposed as model B (\ref{caseB}). The parameters are $T_L=2$, $T_R=T_L-\Delta T$, $\gamma_L=\gamma_R=1$, $\mu_R=0$, $k_R=1$ and $k_L=0.05$.}
\end{figure}
\end{center}

\begin{center}
\begin{figure}[htbp]
\includegraphics[width=0.8\textwidth]{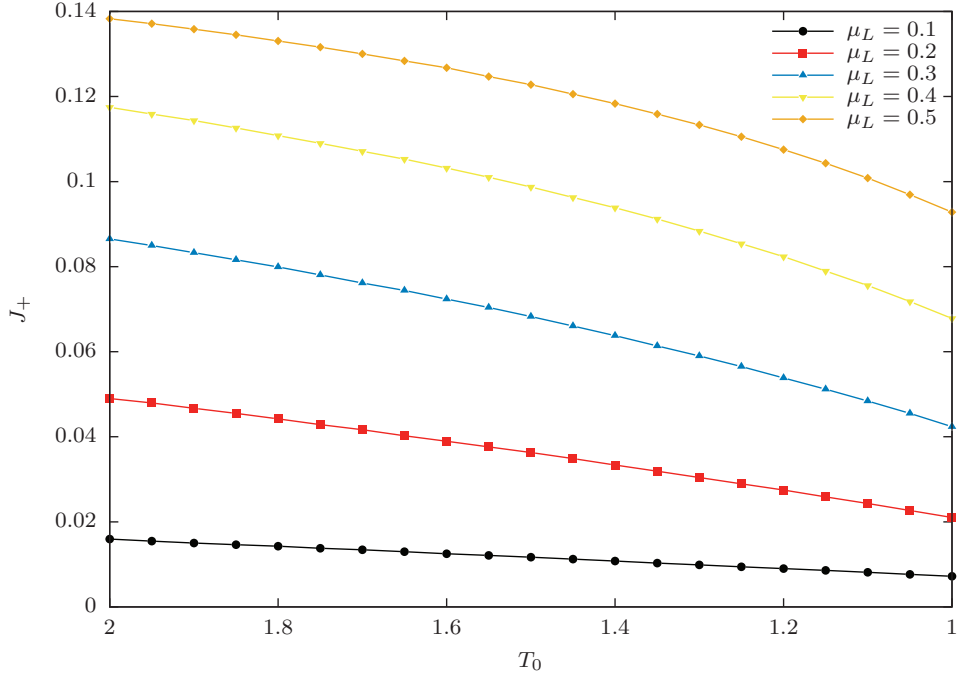}
\caption{\label{fig5} Heat current $J_{+}$ as function of the average temperature $T_0$. The temperature difference is imposed as model A (\ref{caseA}). The parameter is $\Delta T=1$. The other parameters are same as those in Fig.~\ref{fig4}.}
\end{figure}
\end{center}

When the temperature difference is imposed as model B (\ref{caseB}), negative differential thermal conductance can be obtained as predicted in Eq.~(\ref{NDTC2}). According to Eq.~(\ref{NDTC2}), we set $\mu_R=0$, $k_R=1$, $k_L=0.05$ and thus $k_R/k_L=20$, the negative differential thermal conductance is obtained in Fig.~\ref{fig4}. As predicted in Eq.~(\ref{NDTC2}), all the onset temperature differences are larger than $T_L/2=1$. In addition, the higher the $\mu_L$ is, the higher the onset temperature difference is. The appearance of negative differential thermal conductance can be attributed to the same mechanism as in Ref.~[\onlinecite{Hu11}]. With the increasing of $\Delta T$, the average temperature $(T_L+T_R)/2$ decreases. When the heat conductivity decreases with the decreasing $(T_L+T_R)/2$, the negative differential thermal conductance may be obtained. This mechanism is verified in Fig.~\ref{fig5}. The heat current $J_+$ decreases with the decreasing $T_0$. The higher the $\mu_L$ is, the faster the decreasing of $J_+$ is, and thus the faster the decreasing of the corresponding $J_+$ in the negative differential thermal conductance region is as shown in Fig.~\ref{fig4}.

\subsection{strong linear coupling constant}

\begin{center}
\begin{figure}[htbp]
\includegraphics[width=0.8\textwidth]{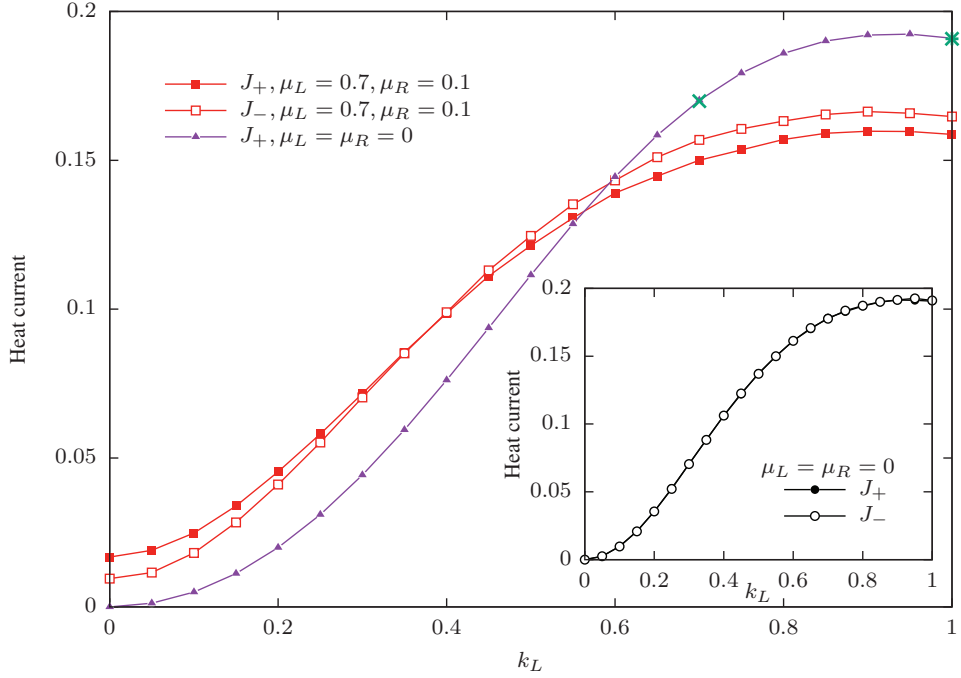}
\caption{\label{fig6} Heat currents $J_{+}$ and $J_-$ as functions of the linear coupling $k_L$. Square symbols and up-triangle symbols correspond to asymmetric nonlinear couplings and symmetric linear couplings respectively.  The temperature difference is imposed as model A (\ref{caseA}). The parameters are $T_0=1.5$, $\Delta T=1$ for $J_+$ and $\Delta T=-1$ for $J_-$, $\gamma_L=\gamma_R=1$ and $k_R=k_L$. The cross symbol and the star symbol correspond to the intercepts $b_1$ and $b_2$ in Fig.~\ref{fig8}. In the inset, $k_R$ is fixed at $1$.}
\end{figure}
\end{center}

With the increasing of the linear coupling constant, more high-frequency phonons can transport across the system-bath interface\cite{Zhang2011}. Therefore, the heat current increases with the linear coupling constant as shown in the inset of Fig.~\ref{fig6}. In addition, when the linear coupling is weak, the heat current is proportional to the square of the linear coupling, which is predicted by $\kappa_0\Delta T$ in Eq.~(\ref{rect}) and is consistent with the results in Refs.~[\onlinecite{Hu2009, Shen2011, Rieder1967}]. (One should note that $\gamma_L k_L^2$ and $\gamma_R k_R^2$ in this work correspond to the friction constant $\lambda$ in Ref. [\onlinecite{Rieder1967}].) Without the nonlinear coupling, thermal rectification is absent as shown in the inset. However, when the asymmetric nonlinear couplings are present, thermal rectification appears in Fig.~\ref{fig6}. Moreover, the direction of thermal rectification reverses as $k_L(=k_R)$ increasing. When the linear coupling is weak, the nonlinear system-bath coupling can enhance the heat current as aforementioned (if $k_L$ is weak, heat currents depicted as square symbols are higher than heat currents depicted as up-triangle symbols in Fig.~\ref{fig6}). The higher nonlinear coupling enables more phonons transport across the interface, and thus the heat current is larger when the hot bath is coupled to the system with higher nonlinear coupling constant. With the increasing of $k_L$ and $k_R$, the linear coupling becomes strong. Consequently, more high-frequency phonons can transport across the interface. Therefore, the umklapp process becomes dominating and thus the nonlinear coupling suppresses the heat current (if $k_L$ is strong, heat currents depicted as square symbols are lower than heat currents depicted as up-triangle symbols in Fig.~\ref{fig6}). The higher the nonlinear coupling is, the more the heat current is suppressed and thus $J_+<J_-$ when $k_L>0.4$. It should be mentioned that even without the nonlinear couplings, the heat current will also be suppressed when the linear couplings are strong enough as shown in Ref. [\onlinecite{Rieder1967}]. However, the mechanism of suppression is potentially the mismatching between frequencies of the bath and the system. Which is different from the mechanism of suppression by the nonlinear couplings.

\begin{center}
\begin{figure}[htbp]
\includegraphics[width=0.8\textwidth]{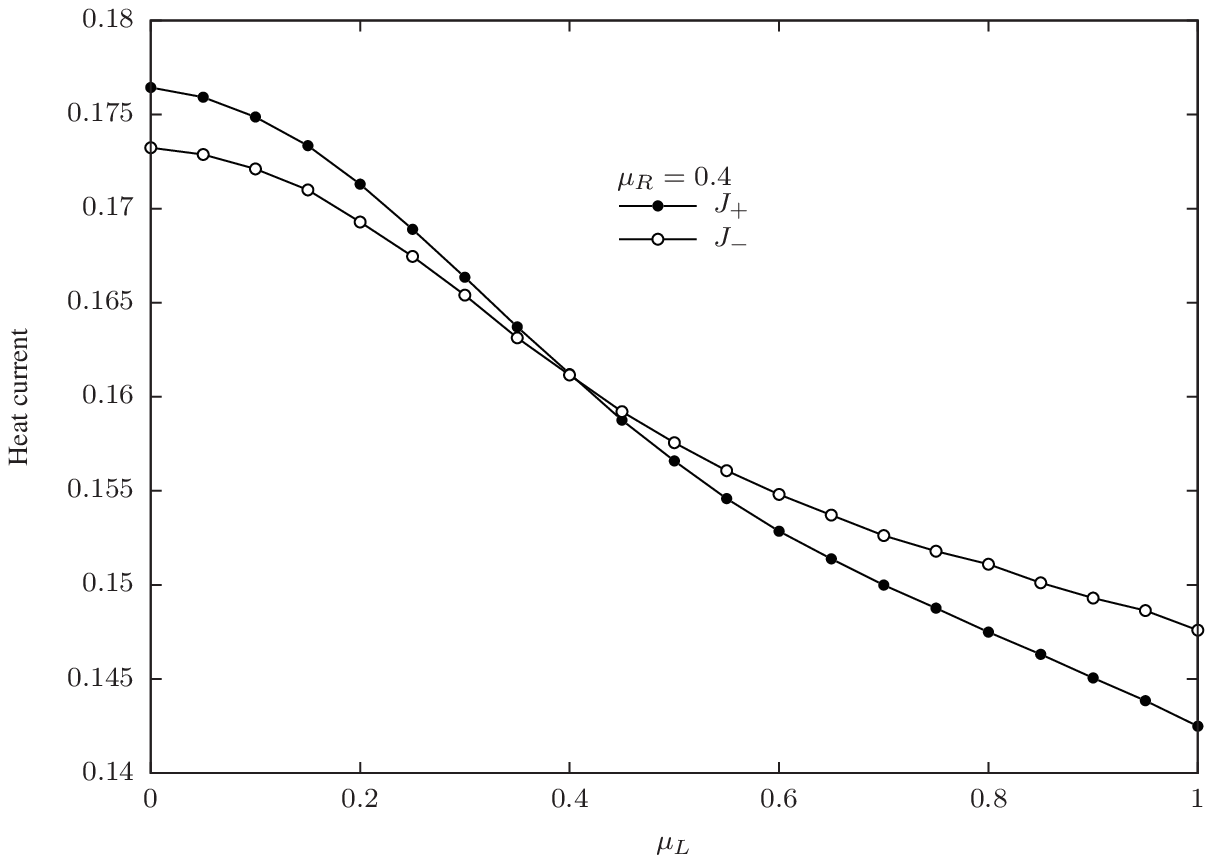}
\caption{\label{fig7} Heat currents $J_{+}$ and $J_-$ as functions of the nonlinear coupling $\mu_L$.  The temperature difference is imposed as model A (\ref{caseA}). The parameters are $T_0=1.5$, $\Delta T=1$ for $J_+$ and $\Delta T=-1$ for $J_-$, $\gamma_L=\gamma_R=1$, $\mu_R=0.4$ and $k_R=k_L=1$.}
\end{figure}
\end{center}

The impacts of nonlinear system-bath coupling on heat current is shown in Fig.~\ref{fig7} when the linear couplings are strong ($k_L=k_R=1$). The heat currents $J_+$ and $J_-$ are suppressed by the nonlinear couplings and thus decrease with $\mu_L$. Higher nonlinear coupling suppresses more heat current. Hence, thermal rectification appears when $\mu_L\neq \mu_R$ and the direction of thermal rectification reverses at $\mu_L=\mu_R$ with the increasing of $\mu_L$.

\begin{center}
\begin{figure}[htbp]
\includegraphics[width=0.8\textwidth]{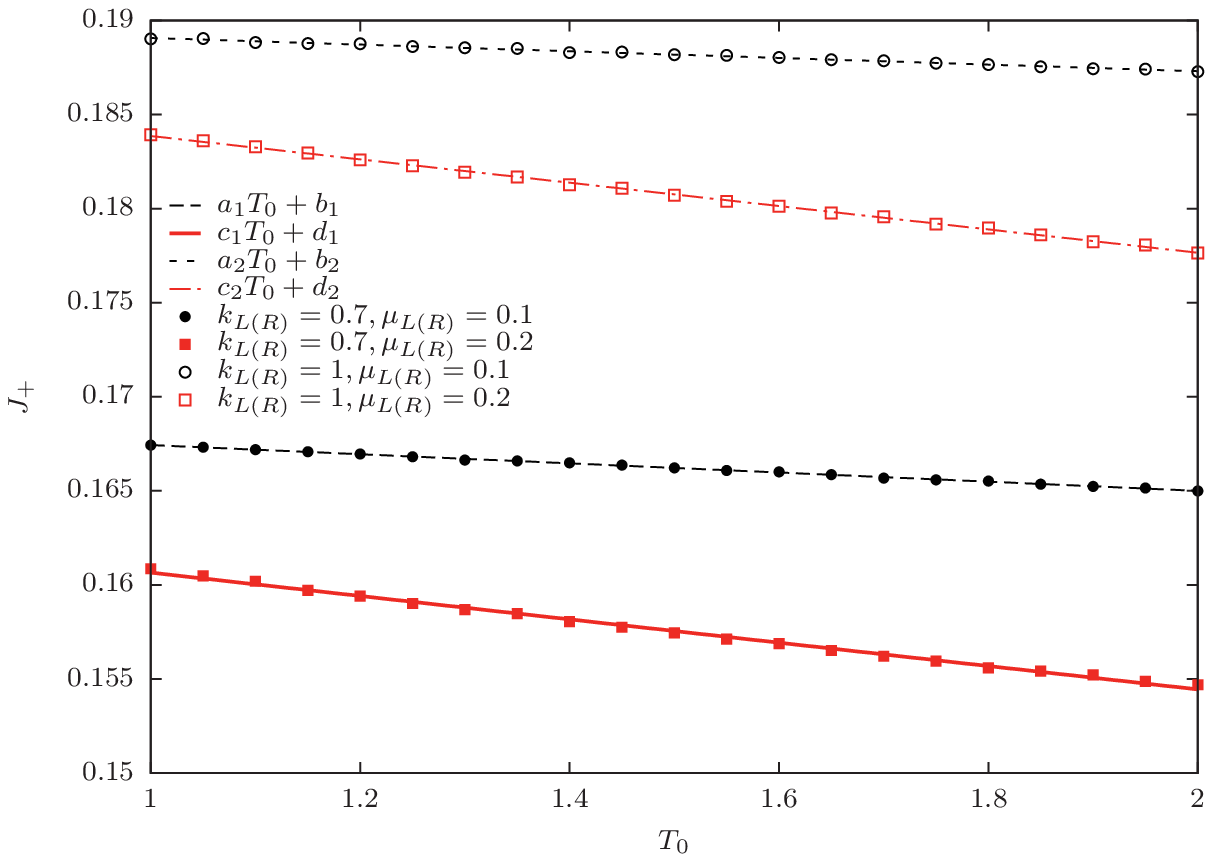}
\caption{\label{fig8} Heat currents $J_{+}$ as functions of the average temperature $T_0$. The temperature difference is imposed as model A (\ref{caseA}). The parameters are $\Delta T=1$ and $\gamma_L=\gamma_R=1$. The circle symbols and the square symbols correspond to $\mu_L=\mu_R=0.1$ and $\mu_L=\mu_R=0.2$. The solid symbols and the open symbols correspond to $k_L=k_R=0.7$ and $k_L=k_R=1$. The long-dash, solid, short-dash and dot-dash lines are linear fits of the data. The fitting parameters are $a_1=-0.00243144$, $b_1=0.169865$, $c_1=-0.00620639$, $d_1=0.16686$, $a_2=-0.00176978$, $b_2=0.190841$, $c_2=-0.00620473$ and $d_2=0.190064$.}
\end{figure}
\end{center}

At high temperature, the population of phonons increases linearly with temperature. Consequently, when the nonlinear coupling is present and the linear coupling constant is strong, there are more high frequency phonons participate the umklapp processes with the increasing temperature. Therefore, the heat conductivity decreases linearly with temperature. As described above, heat current $J_+$ will decrease linearly with the average temperature $T_0$. This is confirmed in Fig.~\ref{fig8}. We obtained that $c_1\approx c_2$. Which means the slopes of the linear fits for $\mu_L=\mu_R=0.2$ are equal. Although $a_1\neq a_2$, the corresponding fitting lines (the long-dash line and the short-dash line) in Fig.~\ref{fig8} approach parallel. We attribute this discrepancy to the fitting errors and the numerical errors. Therefore, the slopes of the linear fits are also $\mu_{L(R)}$ dependent when $k_L=k_R$. The higher the $\mu_{L(R)}$ is, the higher the slope is. This is coincide with the results of weak linear coupling. Furthermore, the intercepts are equal, i.e., $b_1\approx d_1$ and $b_2\approx d_2$. According to the results of weak linear coupling, these intercepts approach the corresponding heat currents of harmonic chain with only the linear system-bath couplings. This is confirmed in Fig.~\ref{fig6}. Where the cross symbol and the star symbol correspond to the intercepts $b_1$ and $b_2$ respectively.

Although the heat conductivity decreases with the temperature when the linear coupling is strong, we have not achieved the negative differential thermal conductance by fixing $T_L$ but increasing $T_R$ (or fixing $T_R$ but increasing $T_L$). We attribute the absence of negative differential thermal conductance to that the slope of decreasing (see Fig.~\ref{fig8}) is much lower than the slope of increasing (see Fig.~\ref{fig3}).

\section{Conclusion and discussion}\label{sec4}
In summary, heat flow in harmonic chain with nonlinear system-bath coupling is studied based on the generalized Caldeira-Leggett model in this work. The obtained Langevin-like equations of motion are solved analytically and numerically. When the linear coupling constant is weak, the numerical results are consistent with the predictions of the approximate analytical results. The heat current is enhanced by the nonlinear system-bath coupling. This is attributed to the fact that the weak linear system-bath coupling allows only the low-frequency phonons to transport across the system-bath interface. When the nonlinear coupling is present, the high-frequency phonons can transport across the interface through the multiphonon processes. Hence the heat current is enhanced. The stronger nonlinear coupling enables more phonons to transport across the interface. Therefore, thermal rectification is obtained when the nonlinear couplings are asymmetric. When both linear couplings are weak, higher heat current is obtained when the hot bath is coupled to the chain with the stronger nonlinear coupling. Moreover, the populations of phonons increase linearly with temperature at high temperature. Therefore, the heat conductivity increases linearly with temperature when the nonlinear system-bath coupling is present. As predicted by the analytical results, by suitable choosing of coupling constants, the negative differential thermal conductance is achieved.

However, when the linear coupling constant is strong, high-frequency phonons can transport across the system-bath interface through linear coupling. The umklapp processes dominate the multiphonon processes when the nonlinear coupling is present. Hence the heat current is suppressed. The stronger nonlinear coupling suppresses more heat current. Therefore, in contrast to the results of weak linear coupling constant, the direction of thermal rectification is reversed, namely, higher heat current is obtained when the hot bath is coupled to the chain with the weaker nonlinear coupling. However, although the heat conductivity decreases with the temperature, the negative differential thermal conductance is not achieved in this work. The potential reason is attributed to the slow decreasing of heat conductivity with temperature.

As stated in Sec.~\ref{sec2-2}, the zeroth approximation is derived for weak nonlinear coupling constant. In addition, the numerical results indicate that the validity of the approximate analytical results is limited to the weak linear coupling constant. It is not valid for the strong linear coupling constant unless the nonlinear coupling constant equals to zero. In deriving the analytical results, Eq.~(\ref{app-eqn3}) is used. It is satisfied only when $T_L=T_R$. However, when the nonlinear coupling constants equal to zero, the analytical results coincide with the reported results in Refs.~[\onlinecite{Dhar2001, Dhar2003, Dhar2006, Dhar2008}]. Therefore, we expect that the analytical results are valid for high temperature difference. This is consistent with the numerical results of weak linear coupling constant.

All the results obtained in this work are based on the nonlinear system-bath coupling. Nonlinear system-bath coupling is non-trivial. The experiments on thermal boundary conductance (interfacial thermal conductance)\cite{Stoner92, *Stoner93, Stevens05, Lyeo06, Hopkins2007, Hopkins2008, Chen2009, Panzer10, Oh2011, Zhang2012, Norris2012, Duda2013, Dechaumphai14, Hohensee2015, Wang2015} reveal that the inelastic phonon scattering at interface between highly dissimilar materials is the dominant reason for the enhancement of heat current. Additionally, at the interface between similar materials, the suppression of heat current relative to the elastic thermal conductance is also observed\cite{Hohensee2015, Stevens05}. Our results of strong linear coupling constant indicate that the nonlinear coupling is one potential reason for the suppression of heat current. Therefore, the nonlinear coupling between different materials is intrinsic. Especially for the nanostructures with high aspect ratio, the nonlinear dissipation is easy achieved. The nonlinear dissipation is significant for nanostructures under tensile stress, but is negligible for them with slack\cite{Eichler2011}. This is consistent with the results of thermal boundary conductance in Ref.~[\onlinecite{Saaskilahti2014}]. In which, under tensile stress, the transmission of phonons through the linear coupling is suppressed but the transmission of inelastic energy is nearly unaffected. This can be understood as follows. The applied tensile stress weakens the linear coupling constant but almost does not impact the nonlinear coupling constant. Therefore, the relative strength of linear coupling and nonlinear coupling can be tuned by applying pressure\cite{Hohensee2015, Saaskilahti2014}.

We hope that our study motivates further research on thermal rectification and negative differential thermal conductance in nanostructures with nonlinear dissipation.

\begin{acknowledgments}
We thank the referees for their constructive comments. Z. J. Ding is supported by the National Natural Science Foundation of China (No. 11274288), the National Basic Research Program of China (Nos. 2011CB932801 and 2012CB933702), Ministry of Education of China (No. 20123402110034) and ``111'' project (No. B07033). Some numerical calculations in this work were performed on the supercomputing system in the Supercomputing Center of University of Science and Technology of China.
\end{acknowledgments}

\appendix

\section{Transformation of Hamiltonian \label{app0}}
 According to Refs.~[\onlinecite{Dhar2003, Dhar2006, Dhar2008}], the Hamiltonian of each bath (\ref{bath}) can be transformed into the normal-mode form by a canonical transformation
\begin{eqnarray}
  Q_\alpha &=& \sum_{s=1}^M U_{\alpha s}\tilde{Q}_s \nonumber\\
  P_\alpha &=& \sum_{s=1}^M U_{\alpha s}\tilde{P}_s.
\end{eqnarray}
Where $U_{\alpha s}$ satisfies
\begin{eqnarray}
&& \sum_{\beta=1}^M K_{\alpha\beta}U_{\beta s}=\omega_s^2U_{\alpha s} \nonumber\\
&& \sum_{\alpha=1}^M U_{\alpha s}U_{\alpha s'}=\delta_{ss'}.
\end{eqnarray}
Hence, Hamiltonian (\ref{bath}) is transformed as
\begin{equation}\label{appa1}
  H_B=\sum_{s=1}^M(\frac{1}{2}\tilde{P}_s^2+\frac{1}{2}\omega_s^2\tilde{Q}_s^2).
\end{equation}
Additionally, the coupling Hamiltonian (\ref{coup}) can be transformed into
\begin{equation}\label{appa2}
  H_I=-\sum_{s=1}^M g(x_1)U_{Ls}\tilde{Q}_s-\sum_{s'=1}^M f(x_N)U_{Rs'}\tilde{Q}_{s'}.
\end{equation}
The transformed Hamiltonians (\ref{appa1}) and (\ref{appa2}) are coincide with them in Refs.~[\onlinecite{Zaitsev12,Barik05}]. Therefore, the generalized Langevin equation (\ref{motion}) can be obtained according to Refs.~[\onlinecite{Zaitsev12,Barik05}].

\section{Entries of matrix $\hat{Z}$ \label{app2}}
As shown in Eq.~(\ref{app-eqn1}), matrix $\hat{Z}$ is just the inversion of matrix $\hat{Y}$. According to Ref.~[\onlinecite{Huang1997}], the entries of matrix $\hat{Z}$ can be calculated as
\begin{eqnarray}\label{appb-eqn1}
  \hat{Z}_{11} &=& \frac{A_{2,N}}{A_{1,N}}, \nonumber\\
  \hat{Z}_{1N} &=& \hat{Z}_{N,1} = \frac{1}{A_{1,N}}, \nonumber\\
  \hat{Z}_{NN} &=& \frac{A_{1,N-1}}{A_{1,N}},
\end{eqnarray}
with
\begin{eqnarray}\label{appb-eqn2}
  A_{1,N} &=& D_{1,N}-\hat{\Gamma}_{11} D_{2,N} -\hat{\Gamma}_{NN} D_{1,N-1} \nonumber\\
  && +\hat{\Gamma}_{11} \hat{\Gamma}_{NN} D_{2,N-1}, \nonumber\\
  A_{1,N-1} &=& D_{1,N-1}-\hat{\Gamma}_{11} D_{2,N-1}, \nonumber\\
  A_{2,N} &=& D_{2,N}-\hat{\Gamma}_{NN} D_{2,N-1}.
\end{eqnarray}
Where $A_{l,m}$ and $D_{l,m}$ are defined as the determinants of the submatrices of $\hat{Y}$ and $\hat{\Phi}-\omega^2 \hat{M}$ beginning with the $l$th row and column and ending with the $m$th row and column. Obviously, $A_{l,m}(-\omega)=A_{l,m}^*(\omega)$ is satisfied, where the star ($^*$) implies the complex conjugate. Therefore,
\begin{eqnarray}\label{appb-eqn3}
  A_{1,N-1}(\omega) A_{1,N-1}(-\omega) &=& D_{1,N-1}^2 +\gamma_L^2 k_L^4 \omega^2 D_{2,N-1}^2 \nonumber \\
  A_{2,N}(\omega) A_{2,N}(-\omega) &=& D_{2,N}^2 +\gamma_R^2 k_R^4 \omega^2 D_{2,N-1}^2,
\end{eqnarray}
with
\begin{eqnarray}\label{appb-eqn4}
  D_{1,N-1} &=& (2-m_1 \omega^2)D_{2,N-1} -D_{3,N-1} \nonumber \\
  D_{2,N} &=& (2-m_N \omega^2) D_{2,N-1}-D_{2,N-2}.
\end{eqnarray}
Hence, when $\gamma_L k_L^2=\gamma_R k_R^2$ and the harmonic chain is equal-mass, one can obtain $\hat{Z}_{11}=\hat{Z}_{NN}$ and then $\hat{I}_{11}=\hat{I}_{NN}$ in Eq.~(\ref{rect}).

\end{document}